\documentclass{article}
\usepackage{microtype}
\usepackage{graphicx}
\usepackage{subcaption}
\usepackage{booktabs} 
\usepackage{hyperref}

\usepackage[preprint]{icml2026}
\makeatletter
\renewcommand{\Notice@String}{}
\makeatother

\usepackage{amsmath}
\usepackage{amssymb}
\usepackage{mathtools}
\usepackage{amsthm}
\usepackage[capitalize,noabbrev]{cleveref}
\usepackage[linesnumbered,ruled,vlined,algo2e]{algorithm2e}
\usepackage{multirow}
\usepackage{booktabs}
\usepackage{caption}
\usepackage{array}

\theoremstyle{plain}

\theoremstyle{definition}

\theoremstyle{remark}

\usepackage{enumitem}
\icmltitlerunning{HeatCache}

\begin{document}

\twocolumn[
\icmltitle{HeatCache: Thermal-aware Energy-efficient LLM Inference Scheduling for Chassis-level Liquid Cooling in Sustainable Edge Server Rooms}
  \begin{icmlauthorlist}
    \icmlauthor{Rui Lu}{polyu}
    \icmlauthor{Huanghuang Liang}{whu}
    \icmlauthor{Kaiqi Guan}{whu}
    \icmlauthor{Dan Wang}{hkust}
  \end{icmlauthorlist}
  \icmlaffiliation{polyu}{Department of Computing, The Hong Kong Polytechnic University, Hong Kong SAR, China}
  \icmlaffiliation{whu}{School of Computer Science, Wuhan University, Wuhan, China}
  \icmlaffiliation{hkust}{Division of Environment and Sustainability, The Hong Kong University of Science and Technology, Hong Kong SAR, China}
  \icmlcorrespondingauthor{Rui Lu}{rui2020.lu@connect.polyu.hk}
  \icmlcorrespondingauthor{Huanghuang Liang}{hhliang@whu.edu.cn}
  \icmlcorrespondingauthor{Kaiqi Guan}{guankaiqi@whu.edu.cn}
  \icmlcorrespondingauthor{Dan Wang}{wangdan@ust.hk}
  \vskip 0.3in
  \icmlkeywords{Machine Learning, ICML}
]
\printAffiliationsAndNotice{}

\begin{abstract}
LLM inference is increasingly deployed at institution-scale edges to meet service requirements. However, multi-GPU inference consumes a large amount of electricity and produces substantial heat. To improve sustainability, operators and regulations often demand raising the ambient setpoint to reduce cooling electricity. This can increase thermal throttling and hardware aging, leading to Service-Level Objective violations.
In this paper, we present \textit{HeatCache}, a thermal-aware, energy-efficient LLM inference scheduler for commercial chassis-level AIO liquid-cooled GPUs at sustainable ambient temperatures.
HeatCache treats AIO loops as a temporary heat buffer, measured by \textit{heat budget} and schedules requests to minimize energy subject to thermal safety and SLO constraints, based on an electrical-informed heat-demand estimation from \textit{HeatiTS}.
We implement HeatCache atop vLLM and show that it reduces computing energy by up to 18.0\%, decreases thermal-throttle exposure by 81.7\%, and maintains SLO violation rates below 0.9\% even up to $48~^{\circ}\mathrm{C}$.
\end{abstract}

\section{Introduction}
\label{sec:intro}
Large language models (LLMs) are rapidly moving from cloud-only to institutional settings like hospitals and universities. Among these, data governance (GDPR, HIPAA, and APPI), service quality, and customization make public APIs unsuitable, necessitating deployment in edge datacenters.
For hardware accessibility, commodity AI servers with multi-GPU capabilities are now comparable to those of mid-scale cloud setups for most tasks \cite{griot2025implementation}.
Together, these trends push compute into small, distributed facilities outside hyperscale clouds, suggesting that private, institution-scale LLM edge server rooms will play a core role in the AI computation infrastructure \cite{technavio2024edgedc, jiang2024megascale, patel2024characterizing}.

Meanwhile, this shift creates acute energy and thermal constraints. Datacenters already consume about 1.5\% of global electricity, and cooling accounts for a substantial share ($\sim 30\%$)~\cite{iea2025energyai, su2019research, wang2024greencooling35}. Compared to hyperscale clouds, edge server rooms typically have less optimized cooling and airflow management, increasing cooling overhead. Dense multi-GPU servers also increase rack power density and intensify local hot spots, making thermal management critical. 
As a result, operators are increasingly running warmer ambient temperature setpoints to reduce energy use, meet sustainability targets, or accommodate limited HVAC controllability \cite{ASHRAE_TC9_9_2021, EUCoC-DC-2024, alrebei2022quantifying}. Prior analyses show that raising 1$~^\circ$C can save 2–5\% of cooling electricity, and up to 4–5\% of total facility power \cite{tropicaldc2023, hriez2025_high_temp, alrebei2022quantifying}. 
However, this directly shrinks thermal headroom under dense multi-GPU inference loads, while cooling operation and LLM serving are typically managed in isolation. The room setpoint is fixed by facility policies, whereas inference schedulers optimize throughput and Service-Level Objectives (SLOs) without considering thermal dynamics \cite{zhang2025thermal, stojkovic2025tapas}.

\begin{figure}[!t]
 
\centering
\includegraphics[trim=0 0 0 0, clip, width=0.985\linewidth]{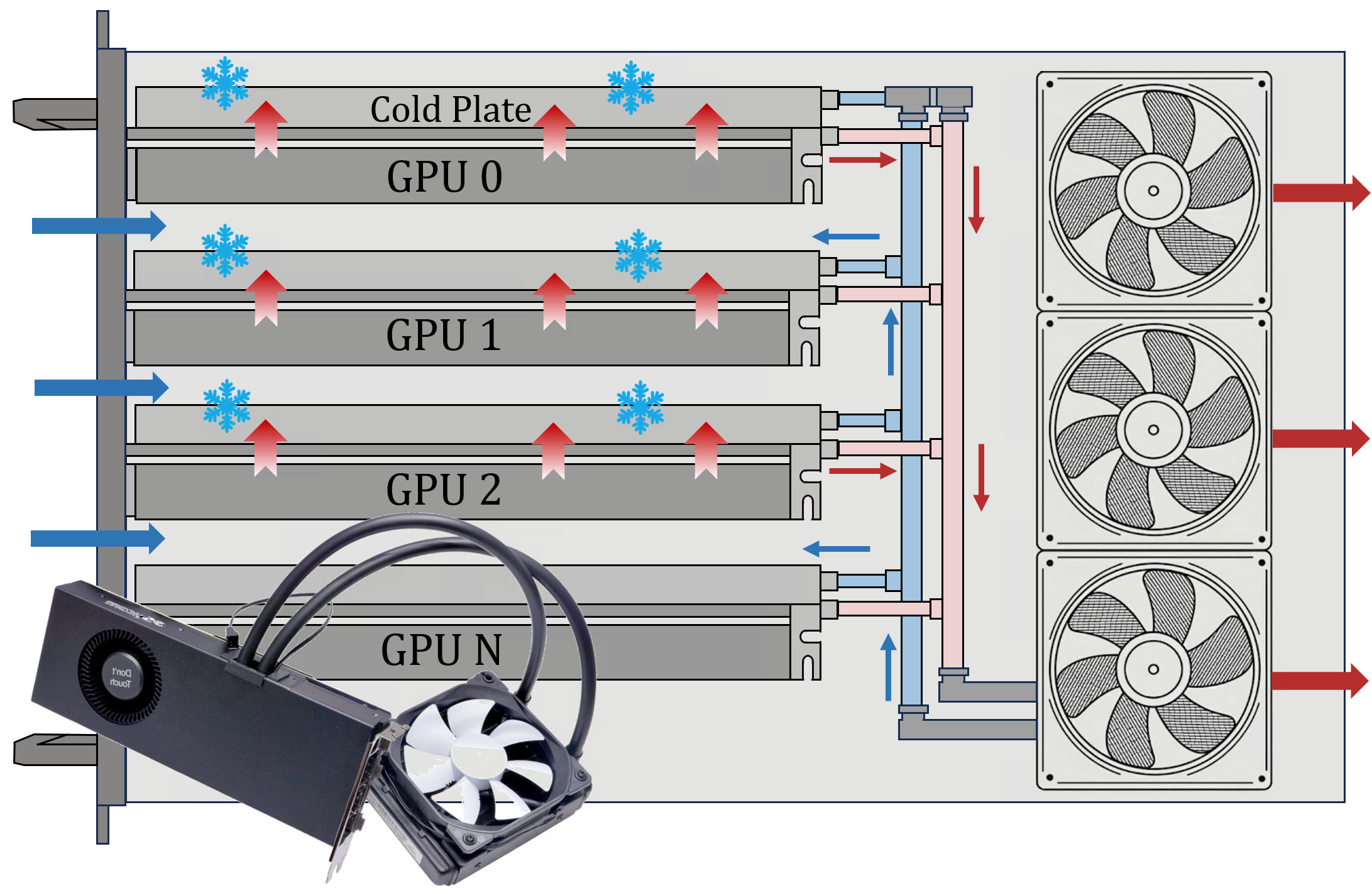}
\caption{
Chassis-level AIO liquid cooling in a server rack.
}\label{fig:AIO}
\end{figure}

This \textit{decoupling} creates a \textit{misalignment} between heat generation and heat dissipation under bursty LLM workloads, causing either unnecessary overcooling or increasing thermal stress, e.g., GPU throttling, and SLO violations when setpoints are raised \cite{kim2021ztt}. These constraints are accelerating liquid cooling adoption as traditional air cooling approaches practical limits under raising GPU power density \cite{wang2024greencooling35}. 
While facility-level liquid cooling can be energy efficient, it typically requires building plumbing, distribution units, and mechanical space that are difficult to retrofit in institutional edge rooms \cite{zhang2023global41degree}. By contrast, chassis-level all-in-one (AIO) water cooling is already available in commercial GPUs, fits existing server chassis and racks, and can be deployed incrementally in small edge server rooms \cite{nvidia_h100_pb11133}. 
Importantly, AIO loops provide nontrivial thermal capacitance in coolant and components, and inference requests can be batched or deferred within tight latency constraints \cite{mittal2014survey}. Together, these properties create an opportunity for a new sight to actively manage heat generation and temporarily buffer heat within the AIO loop, enabling higher room temperatures and lower cooling power without sacrificing thermal safety or SLOs.

In this paper, we study LLM inference on an edge server equipped with chassis-level AIO water-cooled GPUs, deployed in a sustainable server room that operates at warm ambient temperature setpoints. In this setting, the AIO loops provide short-term thermal capacitance that can temporarily absorb additional heat, enabling higher setpoints and lower cooling power. However, aggressive energy savings reduce thermal headroom and can push GPUs into thermal throttling, directly degrading latency and triggering SLO violations \cite{kim2021ztt, stojkovic2025tapas}. Addressing this problem introduces three key challenges.

First, we must quantify how much additional heat the AIO loop can safely buffer before the GPU reaches thermal limits. We call this per-GPU safe buffering capacity the \emph{Heat Budget}.
Second, we must predict the heat that will be generated under different scheduling decisions. We call this predicted thermal load the \emph{Heat Demand}.
Third, we need an online scheduler that matches the forecasted Heat Demand to the available Heat Budget, while jointly respecting SLO constraints and thermal safety.

To address these challenges, we design \emph{HeatCache}, a thermal-aware LLM inference scheduler for sustainable AIO liquid-cooled edge servers. HeatCache aims to minimize total electrical energy while preserving service quality, by guaranteeing SLOs and enforcing per-GPU temperature limits under raising ambient operation. HeatCache consists of three tightly coupled components aligned with the challenges above:
(i) a \emph{heat budget model} that estimates how much additional heat each AIO loop can safely absorb at the current setpoint and operating condition;
(ii) a \emph{workload heat-demand model} that predicts the heat generated by candidate scheduling actions over the job’s execution window; and
(iii) an \emph{online controller} that admits and dispatches batches by matching predicted demand to available budget, enforcing SLO and thermal constraints during runtime. 
Our evaluation shows that HeatCache reduces computing energy by up to 18.0\%, decreases thermal-throttle exposure by 81.7\%, and maintains SLO violation rates below 0.9\% even up to $48~^{\circ}\mathrm{C}$. 
Our contributions can be summarized as:
\begin{itemize} [itemsep=0pt,parsep=0pt,topsep=0pt,itemindent=0pt,labelsep=5pt,leftmargin=10pt]
  \item We first introduce a reduced-order lumped RC model for chassis-level AIO loops. It converts GPU power and ambient conditions into a per-GPU heat cache budget.
  \item We propose HeatiTS, a neural-network job-level power/heat forecaster that maps candidate scheduling actions to predicted heat demand, with a novel electricity regularizer.
  \item We develop a thermal-aware energy-efficient SLO-aware controller to re-rank candidate actions using heat budgets and heat demand.
  \item We design an AIO-cooled testbed for GPU, implement HeatCache on an edge server, and evaluate it with trace-driven LLM inference at raised ambient setpoints. HeatCache reduces energy while preserving SLOs and avoiding throttling.
\end{itemize}
\section{Background and Motivation}
\label{sec:background}

\subsection{Heat and Cooling Challenges in Edge LLM Serving}
\label{sec:bg-hardware-heat}
In GPU-accelerated edge servers, high TDP (300–700 W) leads to significant heat accumulation. During sustained LLM inference, GPU temperatures rise rapidly as power closely tracks utilization. Uneven thermal profiles across GPUs, caused by dynamic routing and batching, often trigger clock-frequency throttling to prevent overheating. This can silently degrade throughput and increase tail latency, leading to SLO violations in latency-sensitive applications. To manage rising power densities, many edge deployments adopt \textit{chassis-level all-in-one (AIO) liquid cooling} (Fig.~\ref{fig:AIO}). Compared to air cooling, AIO cold plates and sealed loops reduce thermal resistance and provide thermal capacitance that can smooth power spikes. However, under high ambient temperatures and sustained load, the loop equilibrates at higher temperatures, reducing thermal headroom and increasing throttling risk. Furthermore, sustainable operation guidelines (e.g., \textit{IEA} and the \textit{EU Code of Conduct}) increasingly encourage higher ambient setpoints and liquid cooling to reduce energy consumption. Consequently, many "sustainable edge server rooms" operate at warmer temperatures, treating thermal configuration as a first-class design knob.

\subsection{LLM Scheduling, SLOs, and Thermal Motivation}
\begin{figure}[t]
    \centering
    \includegraphics[width=0.875\linewidth]{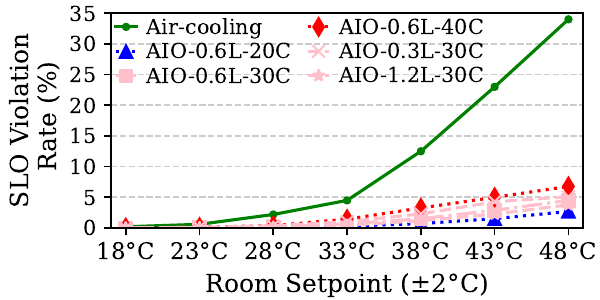}
    \caption{Impact of ambient temperatures and cooling configurations on LLM serving SLO violations.}
    \label{fig:mov-setpoint}
    
\end{figure}

Edge LLM inference relies on batching and routing to meet SLOs (TTFT and TPOT). However, existing schedulers are largely thermal-agnostic, leaving temperature management to hardware. In warm, AIO-cooled environments, scheduling directly impacts heat distribution and throttling. We conducted motivational experiments using Llama-3.1 on an RTX4090 to quantify how ambient setpoints and AIO configurations (coolant volume/temperature) affect TTFT violations. As shown in Fig.~\ref{fig:mov-setpoint}, higher ambient temperatures surge TTFT violations under air cooling (up to ~34\%), whereas AIO cooling maintains stability within a few percent. While warmer coolant increases violations, larger coolant volumes provide thermal buffering to reduce them. These results prove that LLM reliability is highly temperature-dependent. While AIO cooling introduces a complex operating space, it offers superior robustness, motivating the need for the thermal-aware scheduling we propose in HeatCache.

\begin{figure*}[!t]
 
\centering
\includegraphics[trim=0 0 0 0, clip, width=0.95 \textwidth]{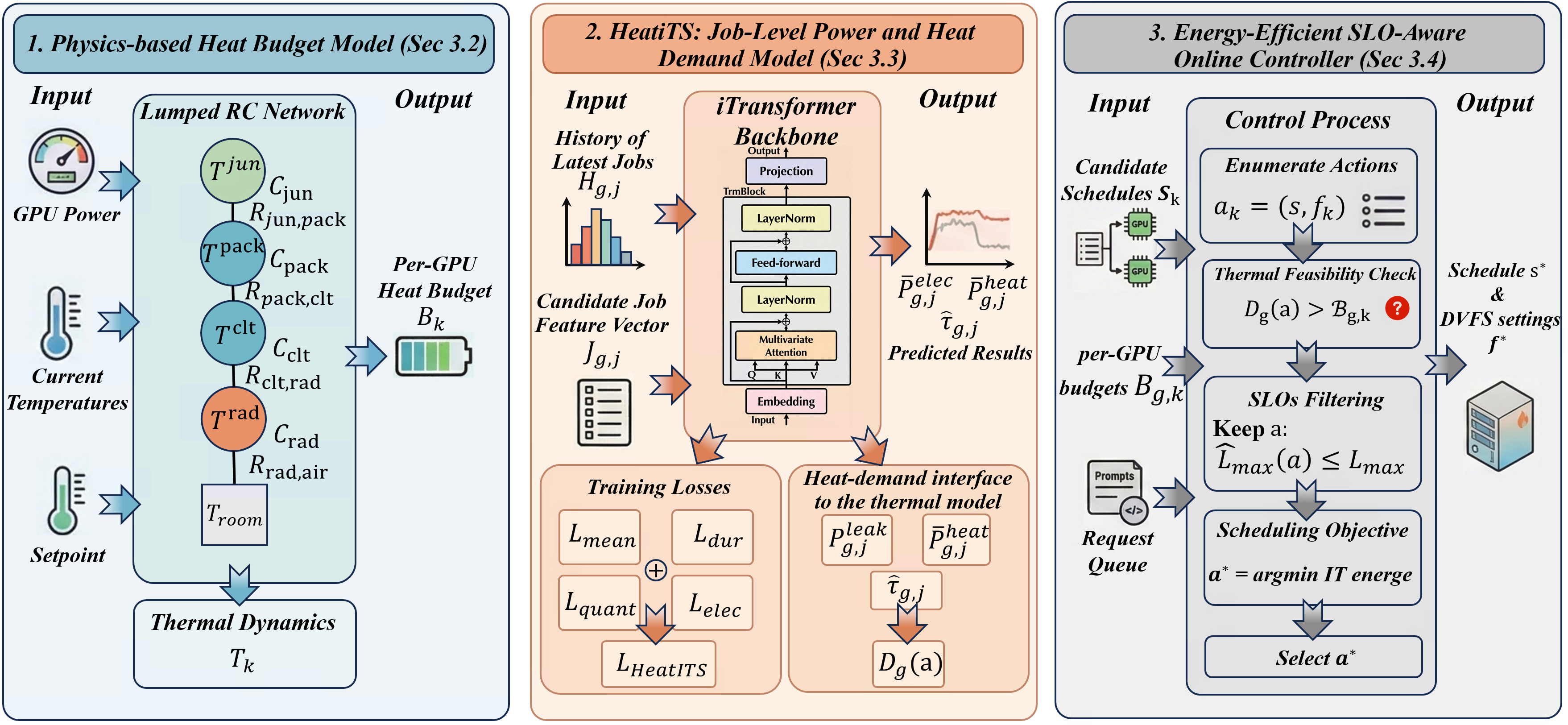}
\caption{
Overview of the HeatCache architecture. 
}
\label{fig:overall_design}
\end{figure*}

\section{Design of HeatCache}\label{sec:design}
\subsection{Design Overview}
\label{sec:design-overview}
HeatCache targets LLM inference on an AIO water-cooled edge server operating in a sustainable server room with a warmer ambient temperature. The objective is to minimize total electrical energy while preserving latency SLOs and avoiding GPU thermal throttling. Achieving this requires coordinating three layers that are typically optimized in isolation: LLM serving decisions (batching and routing), GPU power management (frequency), and thermal behavior of the AIO cooling loops. To structure this coordination, HeatCache is built around three tightly coupled components:
\begin{itemize} [itemsep=0pt,parsep=0pt,topsep=0pt,itemindent=0pt,labelsep=5pt,leftmargin=10pt]
  \item \textbf{Physics-based Heat Budget Model.}
  HeatCache maintains a reduced-order, RC-style thermal model for each GPU and its associated AIO loop. The model is driven by ambient conditions and recent telemetry, providing a fast and physically grounded way to measure the amount of heat that can be absorbed and stored in the water loop.

  \item \textbf{Job-level Power and Heat Demand Model.}
  HeatCache learns a job-level workload model, HeatiTS, that maps queued-request features and candidate serving configurations to heat and power outcomes with risk-aware electrical-regularized training. This allows the controller to anticipate SLOs and the resulting per-GPU power traces induced by each candidate decision.

  \item \textbf{Energy-Efficient SLO-Aware Online Controller.}
  At runtime, HeatCache is designed atop the existing LLM serving stack as a lightweight control layer. At each control interval, it proposes a small set of candidate actions, predicts their latency and power outcomes from HeatiTS, and evaluates their thermal feasibility. It then selects the action that best trades off energy against the enforced SLO and thermal constraints.
\end{itemize}

\paragraph{Design Space and Scheduling Objectives.}
HeatCache targets a single multi-GPU edge server in a sustainable server room with a warmer temperature setpoint $T_{\mathrm{room}}$ for LLM inference tasks. 
An inference \textit{request} $r=\{b^{\mathrm{pf}}_{r},\ell_{r}, m_r\}$ from users stays in a queue $\mathcal{R}$, where $b^{\mathrm{pf}}_{r}$ is the input prompt length for prefill, and $\ell_{r}$ is the prompt-length bucket (e.g., short/medium/long),
and $m_r$ represents the inference LLM (e.g., Llama).
Let $g \in \mathcal{G}$ index an individual GPU. Then a \textit{job} $j\in \mathcal{J}$ is distributed to GPU $g$ to execute one batch of requests $\mathcal{R}_{g,j}\subseteq \mathcal{R}$.

Before $j$ starts, an \textit{action} $a_{g,j}=\{b^{\mathrm{bs}}_{g,j},b^{\mathrm{pf}}_{g,j},f_{g,j}\}$ in all candidates $\mathcal{A}_{g,j}$ taken by the controller specifies the execution details for $g$, including assigned workload amount, i.e., batch size $b^{\mathrm{bs}}_{g,j}$, unified prompt length $b^{\mathrm{pf}}_{g,j} = \max_{r\in \mathcal{R}_{g,j}} b^{\mathrm{pf}}_{r}$, and a DVFS frequency level $f_{g,j}$.
Meanwhile, an \textit{observation} $o_{g,j}=\{\rho_{g,j},P_{g,j},T_{g,j}^{\mathrm{jun}}\}$ taken from GPU telemetry, including on-device sensors and runtime counters, denoting average utilization, average board power, and average junction temperature, respectively, over the execution window.

Each GPU $g$ takes $\Delta t_{g,j}$ to accomplish $j$, and the objective of HeatCache is to minimize the overall IT energy:
\begin{equation}
\min_{\{a_{g,j}\}} \sum_{g\in\mathcal{G}}\sum_{j\in\mathcal{J}} P_{g,j}(a_{g,j})\,\Delta t_{g,j}(a_{g,j}),
\label{eq:oo}
\end{equation}
It minimizes only IT energy, but simultaneously optimizes total facility efficiency under a fixed warm temperature. By constraining thermal safety ($T_{g,j}^{\mathrm{jun}}<T_{\text{thr}}$) and SLO requirements (e.g., TTFT, TPOT), the controller enables the datacenter to maintain a higher ambient setpoint $T_{\mathrm{room}}$, which significantly reduces the cooling overhead.

\subsection{Physics-based Heat Budget Model}
\label{sec:design-thermal}
We model the associated AIO water-cooling loop of each GPU as a \emph{lumped RC thermal network} that captures how heat flows from the GPU die to room air under a given ambient setpoint. The model is fast and physically grounded, so it tracks the thermal state under the measured GPU power and outputs a per-GPU \emph{heat budget}, i.e., the remaining amount of heat the AIO loop can store before the junction reaches the throttling threshold. 

\paragraph{Lumped RC Network.}
Fig.~\ref{fig:rc-model} illustrates the reduced-order thermal circuit for one AIO-cooled GPU.
We use four thermal nodes, including junction, package/cold-plate interface, coolant, radiator core, and treat the room air as a boundary at setpoint $T_{\mathrm{room}}$. Each node has a thermal capacitance, and each link has a thermal resistance. At the endpoint of the link, the radiator-to-air resistance depends on airflow and fan speed. Here, we move the continuous-time RC equations and the state-space matrices to Appendix~\ref{app:rc-details}.

\paragraph{Discrete Thermal Dynamics.}
We discretize the RC network with sampling period $\Delta t$ and obtain the linear update
\begin{equation}
\mathbf{T}_{g,j+1} = A_d \mathbf{T}_{g,j} + B_d P_{g,j} + E_d T_{\mathrm{room}},
\label{eq:rc-discrete}
\end{equation}
where $\mathbf{T}_{g,j} = [T_{g,j}^{\mathrm{jun}}, T_{g,j}^{\mathrm{pack}}, T_{g,j}^{\mathrm{clt}}, T_{g,j}^{\mathrm{rad}}]^\top$. We pre-compute $(A_d,B_d,E_d)$ once per cooling configuration, while the resistances and capacitances are instantiated for a concrete server via calibration in Section~\ref{sec:implementation}. 

\paragraph{Heat Budget as Storable Thermal Capacity.}
Let $T_{\mathrm{thr}}$ be the junction throttling threshold. We first define a \emph{threshold temperature profile} $\mathbf{T}^{\mathrm{thr}}$ across the RC nodes that represents the system state at the moment of throttling:
\begin{equation}
\mathbf{T}^{\mathrm{thr}}(T_{\mathrm{room}})
  = T_{\mathrm{room}}\mathbf{1}
    + (T_{\mathrm{thr}}-T_{\mathrm{room}})\,\mathbf{s},
\label{eq:thr-profile}
\end{equation}
where the \emph{thermal resistance divider ratio} $\mathbf{s}=[1, s^{\mathrm{pack}}, s^{\mathrm{clt}}, s^{\mathrm{rad}}]^\top$ encodes how temperature rise is distributed across the package, coolant, and radiator when the junction is at $T_{\mathrm{thr}}$, see Appendix~\ref{app:par-thermak-res}.

We then define the \emph{heat budget} of GPU $g$, $\mathcal{B}_{g,j}$, as the remaining thermal energy that can be stored in the RC masses before reaching throttling:
\begin{equation}
\mathcal{B}_{g,j}
  = \bigl[\mathbf{C}^\top \bigl(\mathbf{T}^{\mathrm{thr}}(T_{\mathrm{room}}) - {\mathbf{T}}_{g,j} \bigr)\bigr]_+,
\label{eq:heat-budget-state}
\end{equation}
where $\mathbf{C} = [C^{\mathrm{jun}}, C^{\mathrm{pack}}, C^{\mathrm{clt}}, C^{\mathrm{rad}}]^\top$ and $[\cdot]_+$ transform negative values to zero.
This metric depends only on the current thermal state and ambient setpoint, effectively quantifying how much additional heat the AIO loop can absorb before throttling becomes unavoidable.

\subsection{HeatiTS: Job-Level Electrical-regularized Power and Heat Demand Model}
\label{sec:design-heat-demand}

As we mentioned above, the thermal model describes how a given power trace translates into the heat budget.
To use it in the scheduler, we need to know, for each LLM batch that might be assigned to a GPU, how much power and heat that batch will generate on that GPU over its execution.
Accordingly, we develop \emph{HeatiTS}, a per-GPU, job-conditioned time-series forecaster built on an iTransformer backbone with {electrical-regularized training}, which predicts the batch-level power profile and then converts it into heat demand for the next scheduling decisions. 
For each GPU $g$, we maintain a short history of the latest $L$ completed jobs: 
\begin{equation}\label{eq:history}
\mathcal{H}_{g,j}=\left\{o_{g,i},\,\mathcal{R}_{g,i},\,b^{\mathrm{dec}}_{g,i}\right\}_{i=j-L}^{j-1},
\end{equation}
where $b^{\text{dec}}_{g,i}$ is the decode token number.
We model LLM serving as a sequence of jobs assigned to GPUs.
Input the history $\mathcal{H}_{g,j}$ in Eq.~\eqref{eq:history} and candidates action $a_{g,j}\in\mathcal{A}_{g,j}$, HeatiTS predicts $\bigl( \bar{P}^{\mathrm{elec}}_{g,j},\,
             \bar{P}^{\mathrm{heat}}_{g,j},\,
             \hat{\tau}_{g,j} \bigr)$, 
where $\bar{P}^{\mathrm{elec}}_{g,j}$ is the predicted mean electrical power,
$\bar{P}^{\mathrm{heat}}_{g,j}$ is a conservative heat-demand estimate trained as a high quantile,
and $\hat{\tau}_{g,j}$ is the predicted execution time.

\paragraph{Job-Conditioned Variate-Centric Backbone.}
Standard Transformers embed time steps, mixing heterogeneous variables (e.g., temperature in ${}^\circ$C and input prompt length in $10^3$) early in the attention mechanism, which is often suboptimal for multi-physics modeling.
Instead, HeatiTS uses an \textit{iTransformer backbone}~\cite{liu2023itransformer} to model the correlations between distinct physical and logical channels.

HeatiTS embeds all sensors, LLM, and configuration channels into a latent space and stacks them into a set of tokens.
A stack of inverted Transformer blocks applies multi-head self-attention across these tokens, and a position-wise feed-forward network models how the history of jobs on GPU $g$ influences the power of the next job.
A small decoder head maps the final token representations to the job-level outputs $\bar{P}^{\mathrm{elec}}_{g,j}$, $\bar{P}^{\mathrm{heat}}_{g,j}$, and $\hat{\tau}_{g,j}$.
This allows the model to learn temporal dynamics for thermal inertia distinct from the discrete dynamics of token processing, while the self-attention mechanism captures multivariate correlations.

\paragraph{Risk-aware Electrical-regularized Training in HeatiTS.}
HeatiTS is trained on trace data at the job granularity level.
For each completed job $j$ on GPU $g$, we log $\mathcal{H}_{g,j}$ as introduced in Eq.~\eqref{eq:history}.
The training objective combines a mean-squared error on the mean power forecast, a loss on duration, and a quantile loss $\ell_q(\cdot)$ on the heat-demand power:
\[
\mathcal{L}_{\mathrm{mean}}
  = \sum
    \bigl(\bar{P}^{\mathrm{elec}}_{g,j} - P^{\mathrm{obs}}_{g,j}\bigr)^2, 
\]
\[
\mathcal{L}_{\mathrm{dur}}
  = \sum
    \bigl(\hat{\tau}_{g,j} - \tau^{\mathrm{obs}}_{g,j}\bigr)^2,
\]
\[
\small
\mathcal{L}_{\mathrm{quant}}
  = \sum
    \ell_q\bigl(\bar{P}^{\mathrm{heat}}_{g,j}, P^{\mathrm{obs}}_{g,j}\bigr), 
\]

\textit{Physics-Informed Electrical Regularizer.}
To encode a weak prior on how power scales with frequency levels, we use a dynamic-power-inspired baseline for each job $j$,
\[
P^{\mathrm{dyn}}_{g,j}
  = P^{\mathrm{leak}}_g(f_{g,j})
    + \alpha_g\, V(f_{g,j})^2 f_{g,j}\, \phi_{g,j},
\]
where $P^{\mathrm{leak}}_g$ is the leakage or idle power at frequency $f_{g,j}$, $\alpha_g$ is an effective switched-capacitance constant for GPU $g$, $V(f_{g,j})$ is the supply voltage at that DVFS state, and $\phi_{g,j}$ is a calibrated activity factor based on the prompt bucket.
All parameters are obtained from offline calibration.
Then the loss of the electrical regularizer is:
\[
\mathcal{L}_{\mathrm{elec}}
  = \sum
    \bigl(\bar{P}^{\mathrm{elec}}_{g,j} - P^{\mathrm{dyn}}_{g,j}\bigr)^2, 
\]
which pushes the electrical power forecasts toward the dynamic-power form with physical knowledge.
The total HeatiTS loss is
\[
\mathcal{L}_{\mathrm{total}}
  = \mathcal{L}_{\mathrm{mean}}
  + \beta\, \mathcal{L}_{\mathrm{quant}}
  + \gamma\, \mathcal{L}_{\mathrm{elec}}
  + \eta\, \mathcal{L}_{\mathrm{dur}},
\]
with $\beta$, $\gamma$, and $\eta$ controlling the strength of the quantile, electrical, and duration terms.

\paragraph{Heat-demand interface to the thermal model.}
Therefore, the incremental heat into the AIO loop, i.e., heat demand, on action $a$, is computed as:
\begin{equation}
\label{eq:heat-demand}
\mathcal{D}_g(a)
  \approx \bar{P}^{\mathrm{heat}}_{g,j} \, \hat{\tau}_{g,j}, 
\end{equation}
And the electrical power $\bar{P}^{\mathrm{elec}}_{g,j}$ is used to estimate the expected IT energy of job $j$.
In this way, HeatiTS supplies job-level heat-demand and power estimates that are directly compatible with the heat cache budget and the following SLO-aware controller.

\subsection{Energy-Efficient SLO-Aware Online Controller}
\label{sec:design-controller}

We design the job-level scheduling controller of HeatCache, based on top of the existing LLM serving scheduler, i.e., vLLM, that already performs continuous batching and GPU routing functions. We re-rank a small set of candidate scheduling options using (i) the per-GPU \emph{heat budget} from Section~\ref{sec:design-thermal} and (ii) HeatiTS' job-level power/time predictions from Section~\ref{sec:design-heat-demand}.
The key idea is to treat each AIO loop as a temporary heat buffer: the controller permits short, high-throughput bursts when the heat budget is available, and adjusts workload or GPU frequency level when the budget is close to exhaustion, so GPUs avoid thermal throttling while the room operates at a raised setpoint.

\paragraph{Thermal Feasibility Validation.}
For each candidate job $j$ on GPU $g$, we obtained heat budget $\mathcal{B}_{g,j}$ and HeatiTS predicts
$(\bar{P}^{\mathrm{elec}}_{g,j},\, \bar{P}^{\mathrm{heat}}_{g,j},\, \hat{\tau}_{g,j})$ and computes heat demand $\mathcal{D}_g(a)$ in Eq.~\ref{eq:heat-demand}. To enforce thermal limits with low online overhead, we use a \textit{two-stage} check:

\textit{Fast Heat Budget Checking:} 
For each candidate action $a$,
any action $a$ with $\exists g:\ \mathcal{D}_g(a) > \mathcal{B}_{g,j}$ is instantly rejected.
This $O(1)$ check is sufficient for most decisions.

\textit{Precise Full RC Verification:}
If a candidate action consumes a significant fraction, e.g., $>90\%$, of the remaining budget, we would verify it by one-step forward simulation on the discrete thermal dynamics (Eq.~\ref{eq:rc-discrete}). We reject the candidate if the predicted junction temperature $T^{\mathrm{jun}}_{g,j+1}$ exceeds $T_{\mathrm{thr}}$.
This keeps the common case fast while still handling tight-budget edge cases reliably.

\paragraph{SLOs Filtering with Latency Bounds.}
For each candidate action $a$, we apply a lightweight SLO conservative feasibility check.
As $a$ determines, for each GPU $g$,  HeatiTS predicts execution times $\hat{\tau}_{g,j}$, we compute for each request $r$ an estimated time $\hat{L}_r(a)$ with waiting time $\hat{W}_r(a)$. Details are shown in Appendix~\ref{app:slo-mapping}. 

We keep $a$ only if $\hat{L}_{\max}(a) \le L_{\max}$, where $\hat{L}_{\max}(a) = \max_{r \in \mathcal{R}_k} \hat{L}_r(a)$ to summarize the queue by the worst-case predicted latency, and $L_{\max}$ denotes the SLO bound, and when multiple SLO metrics are used, we apply~\eqref{eq:slo-constraints} to each metric and discard any violated actions.

\paragraph{Scheduling Objective.}
Let $\mathcal{A}^{\mathrm{feas}}$ denote the subset of candidates that satisfy both the thermal checks above and the standard latency SLOs (e.g.,  TTFT, TPOT). The controller selects the action $a^\star$ that minimizes the expected IT energy:
\begin{equation}
    a^\star=\arg\min_{a\in\mathcal{A}^{\mathrm{feas}}}
\sum_{{g,j}}\bar{P}^{\mathrm{elec}}_{g,j}\,\hat{\tau}_{g,j}.
\label{eq:optimization-objective}
\end{equation}

\begin{figure}[t]
    \centering
    \includegraphics[width=0.875\linewidth]{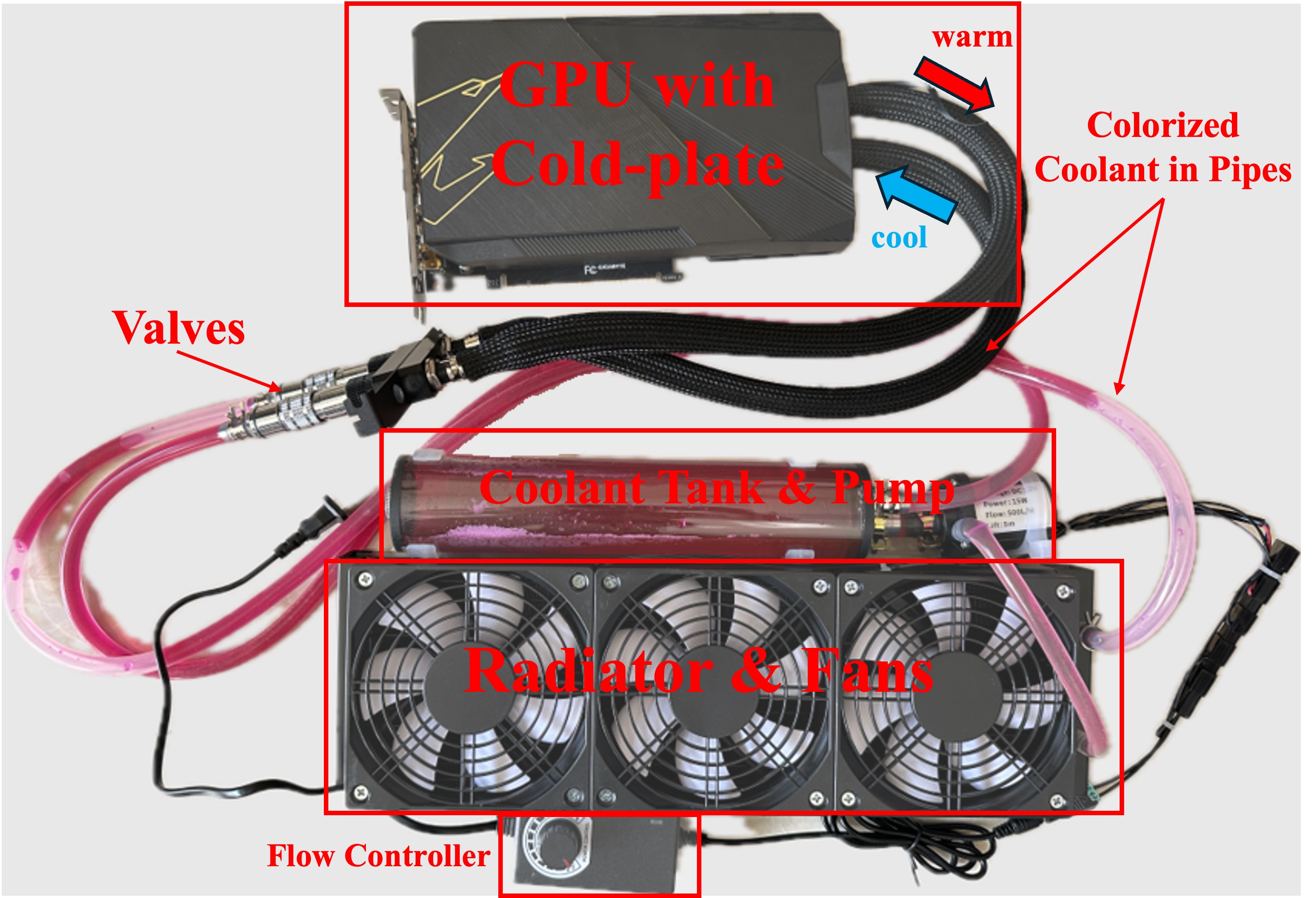}
    \caption{Testbed platform of AIO water-loop GPUs, modified from \textit{AORUS XTREME WATERFORCE} series for investigating distinct existing commercial cooling setups (coolant volume, coolant flow Speed, etc.), with colorized coolant (water). }
    \label{fig:testbed}
    
\end{figure}

\section{Implementation and Evaluation}
\subsection{Implementation}
\label{sec:implementation}
We implement HeatCache as a lightweight user-space plugin on top of the \texttt{vLLM} inference scheduler.
Specifically, HeatCache comprises three modules: 
(i) a physics-based \textit{heat budget monitor}, 
(ii) a job-level \textit{heat demand forecaster} (\emph{HeatiTS}), and 
(iii) an \textit{SLO-aware controller}. 
Our prototype adopts common open-source serving stacks, using \texttt{Kubernetes} for cluster management and \texttt{KServe} as the inference interface. 
We obtain GPU telemetry and frequency control via \texttt{NVIDIA DCGM} and a lightweight \texttt{Kubelet} plugin. 
At runtime, HeatCache adjusts \texttt{vLLM} to tune micro-batch sizes online, and extends the \texttt{KServe} router to parse per-request SLO metadata while preserving fair routing under dynamic batching. 
To ensure safe actuation, HeatCache restricts DVFS actions to a predefined whitelist of stable frequency states and resets devices to a baseline state upon exit or error. 
Finally, we instantiate a per-GPU lumped RC thermal model and calibrate its parameters offline using step-load experiments, see Appendix~\ref{app:rc-calibration}.

\noindent\textbf{AIO Liquid-cooling Testbeds.} 
Our testbed emulates GPUs equipped with chassis-level AIO liquid-cooling loops. See Fig.~\ref{fig:testbed}, a cold plate is mounted on the GPU to extract heat, while a closed coolant loop circulates fluid (600 mL as default) between the GPU junction and an external heat exchanger. Two coolant pipes form the supply (cool) and return (warm) paths, and inline quick-disconnect valves enable safe isolation, filling, and maintenance. The warmed coolant is pumped through an external radiator–fan module to dissipate heat to ambient air. A pump is connected to a coolant tank, stabilizing the coolant volume, reducing bubble formation, and maintaining circulation.  
A flow controller adjusts the coolant rate to tune heat removal and buffering.
The pump and fans consume about 25\,W per loop on average.
We exclude this 25\,W auxiliary power when comparing GPU computing energy, but include it in total system energy. 
This modular setup enables controlled measurements of GPU thermal dynamics and cooling-loop responses under LLM inference workloads and different ambient setpoints.

\begin{figure*}[t]
    \centering
    \begin{subfigure}[t]{0.245\linewidth}
        \centering
        \includegraphics[width=\linewidth]{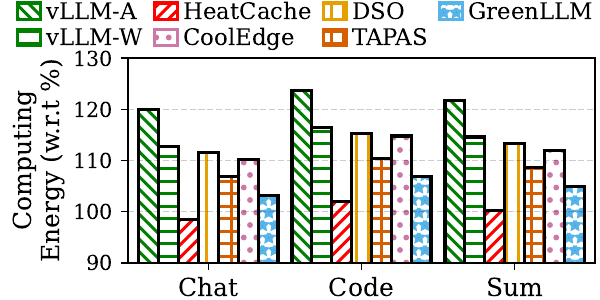}
        \caption{Relative Computing Energy $\downarrow$}
        \label{fig:sub:eva_comp}
    \end{subfigure}
    \hfill
    \begin{subfigure}[t]{0.245\linewidth}
        \centering
        \includegraphics[width=\linewidth]{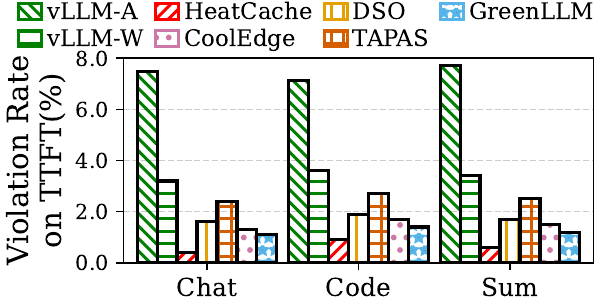}
        \caption{SLO Violation on TTFT $\downarrow$}
        \label{fig:sub:eva_ttft}
    \end{subfigure}
    \hfill
        \begin{subfigure}[t]{0.245\linewidth}
        \centering
        \includegraphics[width=\linewidth]{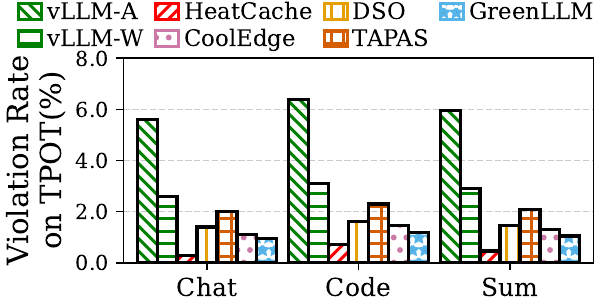}
        \caption{SLO Violation on TPOT $\downarrow$}
        \label{fig:sub:eva_topt}
    \end{subfigure}
    \hfill
    \begin{subfigure}[t]{0.245\linewidth}
        \centering
        \includegraphics[width=\linewidth]{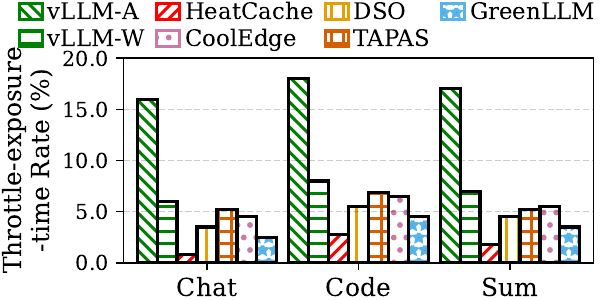}
        \caption{Thermal Safety $\downarrow$}
        \label{fig:sub:eva_tet}
    \end{subfigure}
    \caption{Experiments under \textit{Chat}, \textit{Code}, \textit{Sum}, at sustainable setpoint (48$\pm2^\circ\mathrm{C}$), where (\ref{fig:sub:eva_topt}) is \textit{with respect to} \texttt{vLLM-A}@18$^\circ\mathrm{C}$.}
    \label{fig:eva-res}

\end{figure*}

\begin{figure*}[t]
    \centering
    \begin{subfigure}[t]{0.245\linewidth}
        \centering
        \includegraphics[width=\linewidth]{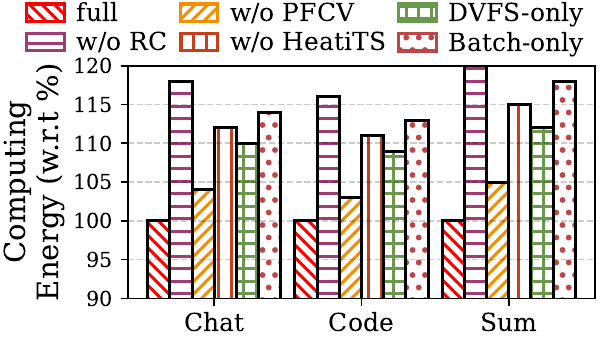}
        \caption{Relative Computing Energy $\downarrow$}
        \label{fig:sub:aba_comp}
    \end{subfigure}
    \hfill
    \begin{subfigure}[t]{0.245\linewidth}
        \centering
        \includegraphics[width=\linewidth]{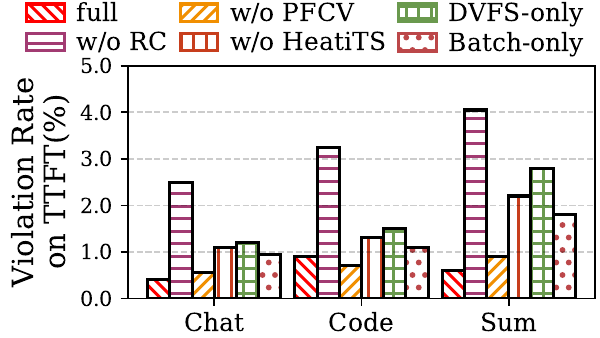}
        \caption{SLO Violation on TTFT $\downarrow$}
        \label{fig:sub:aba_ttft}
    \end{subfigure}
    \hfill
        \begin{subfigure}[t]{0.245\linewidth}
        \centering
        \includegraphics[width=\linewidth]{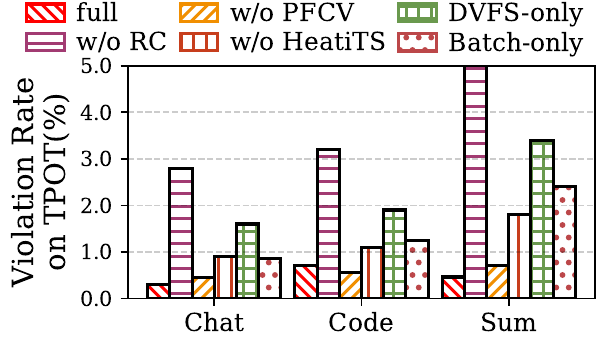}
        \caption{SLO Violation on TPOT $\downarrow$}
        \label{fig:sub:aba_topt}
    \end{subfigure}
    \hfill
    \begin{subfigure}[t]{0.245\linewidth}
        \centering
        \includegraphics[width=\linewidth]{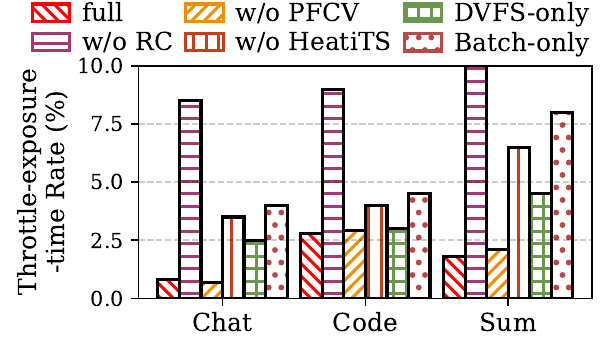}
        \caption{Thermal Safety $\downarrow$}
        \label{fig:sub:aba_tet}
    \end{subfigure}
    \caption{Ablation results under \textit{Chat}, \textit{Code}, \textit{Sum}, at sustainable setpoint (48$\pm2^\circ\mathrm{C}$), where (\ref{fig:sub:aba_comp}) is \textit{with respect to} \textit{full}.}
    \label{fig:aba-res}

\end{figure*}

\subsection{Evaluation Setup}
\label{sec:eval_setup}
We evaluate HeatCache in terms of energy efficiency, SLOs, and thermal safety under diverse workloads and sustained setpoints. Our goal is to quantify the end-to-end benefits of HeatCache while operating within realistic facility constraints and practical LLM serving stacks.

\noindent\textbf{Testbeds.} We validate HeatCache on a workstation server, with an Intel Core i9-13900K CPU and 128 GB RAM. The system features four RTX 4090 GPUs with 24 GB of VRAM, equipped with water-cooling modifications as illustrated in Fig.~\ref{fig:testbed}.
To ensure the models fit within GPU memory, we quantize the served LLMs accordingly. 
The workstation is placed inside a $2\,\mathrm{m}\times2\,\mathrm{m}$ enclosure to emulate a compact edge server room, and the controller selects a continuous temperature setpoint within $18^\circ\mathrm{C}$--$48^\circ\mathrm{C}$. 

\noindent\textbf{LLM inference workloads.}
We evaluate \textit{Llama} under three open-source prompt traces that represent common LLM serving behaviors
\textit{Chat}: we use the \textit{LMSYS-Chat-1M} real-world conversation dataset collected from Chatbots~\cite{zheng2023lmsys}.
\textit{Code}: we use \textit{CodeAlpaca-20K} to emulate developer-style programming requests~\cite{codealpaca}.
\textit{Sum}: we use \textit{QReCC}, an open-domain conversational information-seeking dataset, to capture retrieval-style and summarization-like queries~\cite{anantha2021open}.
Each category reflects distinct token characteristics with specific SLOs definition for each workload, as detailed in Table~\ref{tab:workloads_slo}.

\begin{table}[t]
\centering

\caption{Workload Profiles and Service Level Objectives}
\label{tab:workloads_slo}
\resizebox{0.975\linewidth}{!}{
\setlength{\tabcolsep}{0.7mm}{
\begin{tabular}{c|ccc}
\toprule
\textbf{Inference Task} & \textbf{Chat} & \textbf{Code} & \textbf{Sum} \\ \midrule
\textbf{Trace} & \textit{LMSYS-Chat-1M} & \textit{CodeAlpaca-20K} & \textit{QReCC} \\
\textbf{LLM} & \textit{Llama-3.1} & \textit{Qwen2.5-Coder} & \textit{DeepSeek-R1} \\
\textbf{Arrival Pattern} & Bursty & Poisson & Poisson \\
\textbf{SLO (TTFT)} & $\leq$800\,ms & $\leq$1500\,ms & $\leq$2000\,ms \\
\textbf{SLO (TPOT)} & $\leq$50\,ms & $\leq$60\,ms & $\leq$80\,ms \\
\bottomrule
\end{tabular}
}}

\end{table}

\noindent\textbf{Baselines.}
We compare HeatCache against representative LLM serving and DVFS-based energy control baselines:
\noindent $~\bullet~$\textbf{vLLM} serves as the default inference scheduler baseline, with both air-cooling testbed \texttt{vLLM-A} and AIO water-cooling testbed \texttt{vLLM-W}.
\noindent $~\bullet~$\textbf{DSO} \cite{wang2024dso} sets GPUs to a fixed frequency to meet SLOs without adapting to workload dynamics or thermal states.
\noindent $~\bullet~$\textbf{TAPAS} \cite{stojkovic2025tapas} is an adapted thermal-aware policy that routes each incoming batch to the currently cooler GPU
based on predicted temperature and power.
\noindent $~\bullet~$\textbf{GreenLLM}~\cite{tian2024greenllm} applies SLO-aware, phase-specific DVFS for prefill and decode, and routes requests using input-length-aware queues to reduce blocking.
\noindent $~\bullet~$\textbf{CoolEdge} \cite{pei2022cooledge} is a cooling-aware controller that reduces liquid-loop energy by dynamically regulating coolant delivery (e.g., water temperature/flow) with customized water loops. We imply the algorithm design only on our AIO water loop, but retain the hardware energy reduction shared during accounting.

\noindent\textbf{Metrics.}
We evaluate the following aspects: 
i) the average per-job GPU computing energy, 
ii) the serving quality measured by SLO violation rates (TTFT and TPOT), 
iii) thermal safety using \textit{throttle exposure time rate}, i.e., the fraction of execution spent above a high-temperature threshold.

\subsection{Evaluation Results}
\noindent\textbf{Improvement of Energy-efficiency.}
We compare the energy performance of HeatCache with existing baselines. All energy metrics are normalized to the \texttt{vLLM-A} $@ 18^\circ\mathrm{C}$ configuration in each task in Fig.~\ref{fig:sub:eva_comp}.
We observe both \texttt{vLLM-A/W} increase computing energy as higher temperatures trigger throttling and increase latency. Thus, their reduction is limited, with at most {6.0\%} over \texttt{vLLM-A}. 
DSO is thermally agnostic, obtaining only {6.8\%--7.0\%} reduction. 
TAPAS routes batches to cooler GPUs, but without modeling AIO thermal inertia, throttle happens. 
GreenLLM combines phase DVFS with length-aware routing, but conservative control extends runtime and reduces power gains. 
CoolEdge targets reducing water-loop energy. 
HeatCache is consistently superior by scheduling with AIO thermal buffering, estimating available headroom and avoiding throttle, reducing computing energy up to {18.0\%}.

\noindent\textbf{Improvement of Service Quality. }
We next evaluate SLO violations for TTFT (Fig.~\ref{fig:sub:eva_ttft}) and TPOT (Fig.~\ref{fig:sub:eva_topt}), where lower indicates better service quality.
HeatCache achieves the lowest violations by considering each GPU's thermal headroom and preventing throttle-induced latency spikes.
Among all, HeatCache reduces SLO violations up to 82.5\% compared to \texttt{vLLM-A/W}, which are unaware of thermal transients and frequently overload already-hot GPUs.
DSO improves request distribution but ignores thermal constraints, so violations remain at 1.4\%--1.9\%.
TAPAS routes without accounting for thermal, so headroom is quickly depleted and violations persist.
GreenLLM suppresses short peaks via DVFS, but conservative control can extend runtime and increase queueing delays, worsening latency.
CoolEdge primarily saves water-loop energy, making it similar to \texttt{vLLM-W}.
By shaping heat generation and routing to available headroom, HeatCache consistently minimizes SLO violations.

\noindent\textbf{Improvement of Thermal Safety.}
We quantify thermal safety using \emph{throttle exposure time} in Fig.~\ref{fig:sub:eva_tet}. Higher exposure accelerates aging and increases shutdown risk.
We see that \texttt{vLLM-A/W} has high exposure due to repeated throttling.
TAPAS routes to cooler GPUs, but burst arrivals can quickly drain headroom and bring throttle.
DSO and GreenLLM reduce power peaks via frequency control and indeed reduce the throttling time.
CoolEdge focuses on loop energy and also reduces throttling time.
HeatCache achieves the lowest exposure (0.8\% to 2.8\%) by controlling heat generation and scheduling within available AIO buffering.

\subsection{Ablation Studies}
We ablate HeatCache by disabling key components and control knobs: 
\textit{full} denotes the complete designs. 
\textit{w/o RC} removes the RC model using a traditional threshold heuristic on junction temperature. 
\textit{w/o HeatiTS} disables the HeatiTS heat-demand forecaster and uses physics-only estimation. 
\textit{DVFS-only} enables frequency control only, whereas \textit{Batch-only} enables micro-batch adaptation but fixes GPU frequency. 
\textit{w/o PFCV} disables Precise Full RC Verification during scheduling. The results are shown in Fig.~\ref{fig:aba-res}.

\noindent\textbf{Impact of Heat Budget Design.}
Compared to \textit{w/o RC}, \textit{full} reduces TTFT/TPOT SLO violations by 80.50\%/86.07\% and throttle exposure by 81.69\%, with the largest gap on \textit{Sum}, where heat accumulates quickly.
Reactive threshold control, \textit{w/o RC}, is overly conservative, leading to higher energy and worse tail latency.
\textit{w/o PFCV} shows comparable throttle exposure to \textit{full}, since it does not recompute full thermal constraints at each decision, but it decreases latency by 7\% compared to \textit{full} because \textit{full} incurs added verification overhead.

\noindent\textbf{Impact of HeatiTS Design.}
HeatiTS provides the job-specific heat demand for stable energy--SLO tradeoffs.
\textit{w/o HeatiTS} relies on physics-only estimates that miss prompt-length and burst-dependent heat dynamics, causing either unsafe underestimation, i.e., more throttling, or conservative overestimation, i.e., underutilized headroom.
As a result, \textit{w/o HeatiTS} has higher SLO violations and higher computing energy than \textit{full}, especially under mixed prompt lengths and bursty arrivals.
By contrast, HeatiTS enables accurate headroom allocation while maintaining thermal safety.

\noindent\textbf{Impact of Controller Design.}
Both control knobs are necessary, but neither is sufficient alone.
\textit{DVFS-only} reduces throttle exposure by 38.19\% compared to \textit{Batch-only}, yet its slower token processing increases queueing and degrades SLOs.
\textit{Batch-only} improves throughput opportunistically, but cannot prevent thermal saturation during sustained bursts, resulting in higher throttle exposure and increased SLO violations.
The \textit{full} controller jointly tunes micro-batching and DVFS, achieving the most consistent computing-energy reduction while preserving SLO compliance and minimizing throttling across all workloads.

\section{Related Works}
\label{sec:relatedworks}
\noindent\textbf{Thermal Management in GPU Computing.}
Thermal constraints challenge GPU workloads, where sustained AI compute causes throttling and degrades performance~\cite{tan2024thermal}. Mitigations include thermal-aware workload migration using heuristics or RL~\cite{tan2024thermal}, and balanced scheduling for CPU–GPU systems under tight thermal limits~\cite{lee2019thermal,lee2021thermal}. While these treat thermal headroom as a resource, they assume air cooling and ignore cooling dynamics and LLM scheduling specifics. HeatCache addresses this by modeling water-cooling loops as thermal buffers and coordinating job heat demand to prevent throttles at high ambient temperatures.

\noindent\textbf{Datacenter Cooling and Energy Efficiency.}
Datacenter cooling optimization has progressed from reactive control to proactive thermal management. Traditional approaches focus on airflow design, CRAC tuning, and higher ambient setpoints that deliver 2–5\% energy savings per degree~\cite{tropicaldc2023,hriez2025_high_temp}. Chassis-level AIO liquid cooling is increasingly deployed for rising rack densities. However, these methods optimize cooling independently, treating workloads as exogenous rather than coordinating placement with thermal dynamics.
As datacenters consume 1.5\% of global electricity~\cite{iea2025energyai}, energy-aware inference techniques have emerged, including workload consolidation~\cite{hu2017coordinating}, carbon-aware scheduling~\cite{wiesner2021let,lin2022carbon}, learning-based meta-schedulers~\cite{cheng2021network,aaen2023automatic}, and GPU right-sizing~\cite{Wang2022Energy,sun2025learning}. Yet these assume instantaneous cooling responses and overlook thermal inertia. At the cluster scale, TAPAS~\cite{stojkovic2025tapas} and TAWS~\cite{lu2025thermal} leverage thermal signals for placement but do not model per-device capacitance. Sustainability guidelines~\cite{iea2025energyai,tropicaldc2023} encourage elevated setpoints and liquid cooling. HeatCache operationalizes this by modeling AIO loop inertia at the edge, coordinating heat demand and budgets to support higher ambient temperatures with lower energy under SLO constraints.

\noindent\textbf{LLM Inference Scheduling and SLO Management.}
Modern LLM systems prioritize throughput and SLOs (TTFT, TPOT), employing optimizations like continuous batching (vLLM~\cite{li2023vllm}), kernel fusion (TensorRT-LLM~\cite{nvidia_tensorrtllm_repo}), and advanced scheduling~\cite{Jiang2025Efficient}. These schedulers treat thermal behavior as external, which fails in warm edge environments where compute generates heat and causes throttling. While energy-aware systems use DVFS~\cite{Wang2022Energy,sun2025learning}, they ignore cooling dynamics. HeatCache addresses this by integrating heat prediction and thermal-aware scheduling to meet SLOs and prevent throttling.

\section{Conclusion}
In this paper, we present HeatCache, a thermal-aware scheduling controller for sustainable institution-scale LLM inference under rising ambient setpoints. 
We show that chassis-level AIO liquid cooling provides transient thermal buffering that can be exploited by the serving stack. HeatCache integrates thermal headroom awareness into batching and GPU assignment decisions, proactively avoiding throttle while maintaining service quality. Our prototype demonstrates that thermal-aware scheduling is a practical lever for reliable, energy-efficient edge LLM serving without infrastructure retrofits.
\bibliography{ref}
\bibliographystyle{icml2026}
\newpage
\appendix
\onecolumn
\section{Lumped RC Thermal Model Details}
\label{app:rc-details}

\subsection{RC network and continuous-time dynamics}
\label{app:rc-ct}

\begin{figure}[h]

\centering
\includegraphics[trim=0 0 0 0, clip, width=0.495 \linewidth]{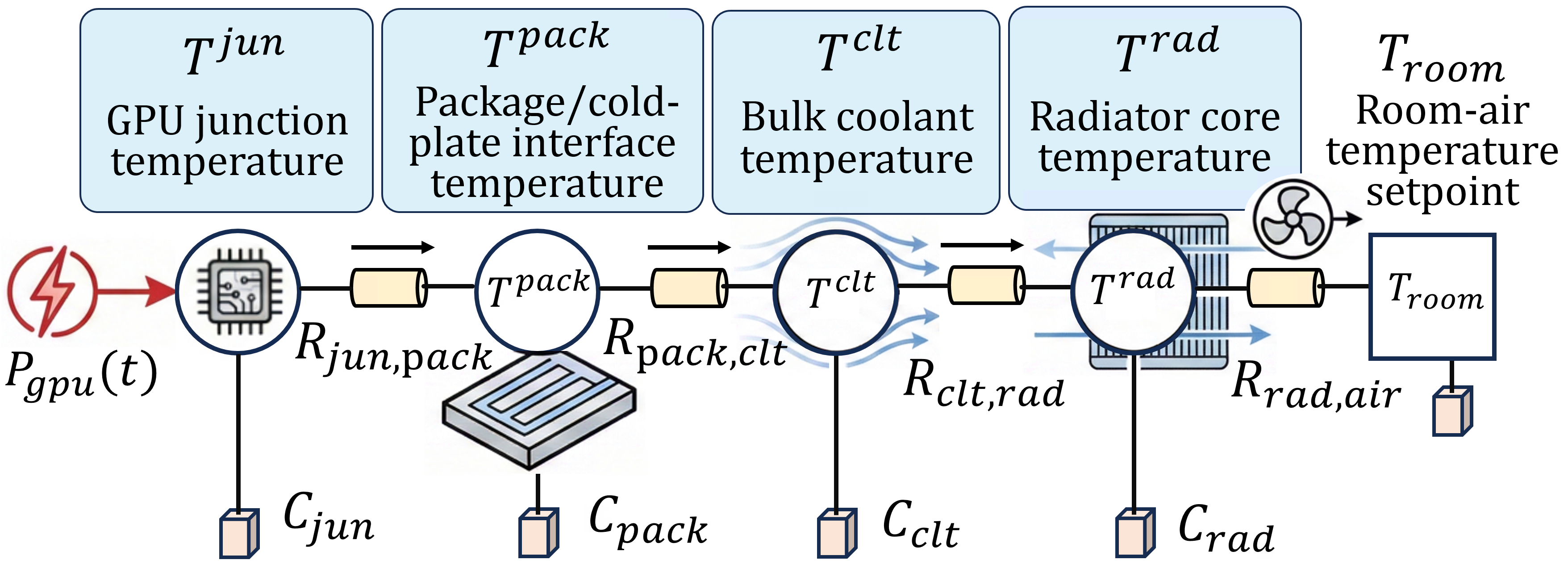}
\caption{
The lumped RC network for a single AIO-cooled GPU.
}
\label{fig:rc-model}
\end{figure}

We use a four-node lumped thermal network as shown in Fig.~\ref{fig:rc-model}. For each GPU $g$, the nodes are:
junction temperature $T^{\mathrm{jun}}_{g}(t)$, package/cold-plate interface temperature $T^{\mathrm{pack}}_{g}(t)$,
bulk coolant temperature $T^{\mathrm{clt}}_{g}(t)$, and radiator-core temperature
$T^{\mathrm{rad}}_{g}(t)$. Room air is an exogenous boundary at setpoint $T_{\mathrm{room}}$.

Energy balance follows:
\begin{align}
C_{\mathrm{jun}} \frac{d T^{\mathrm{jun}}_{g}}{dt}
  &= P_{g}(t)
     - \frac{T^{\mathrm{jun}}_{g}(t) - T^{\mathrm{pack}}_{g}(t)}{R_{\mathrm{jun,pack}}}, \label{eq:rc-j}\\[0.3em]
C_{\mathrm{pack}} \frac{d T^{\mathrm{pack}}_{g}}{dt}
  &= \frac{T^{\mathrm{jun}}_{g}(t) - T^{\mathrm{pack}}_{g}(t)}{R_{\mathrm{jun,pack}}}
     - \frac{T^{\mathrm{pack}}_{g}(t) - T^{\mathrm{clt}}_{g}(t)}{R_{\mathrm{pack,clt}}}, \label{eq:rc-p}\\[0.3em]
C_{\mathrm{clt}} \frac{d T^{\mathrm{clt}}_{g}}{dt}
  &= \frac{T^{\mathrm{pack}}_{g}(t) - T^{\mathrm{clt}}_{g}(t)}{R_{\mathrm{pack,clt}}}
     - \frac{T^{\mathrm{clt}}_{g}(t) - T^{\mathrm{rad}}_{g}(t)}{R_{\mathrm{clt,rad}}}, \label{eq:rc-c}\\[0.3em]
C_{\mathrm{rad}} \frac{d T^{\mathrm{rad}}_{g}}{dt}
  &= \frac{T^{\mathrm{clt}}_{g}(t) - T^{\mathrm{rad}}_{g}(t)}{R_{\mathrm{clt,rad}}}
     - \frac{T^{\mathrm{rad}}_{g}(t) - T_{\mathrm{room}}}{R_{\mathrm{rad,air}}}. \label{eq:rc-r}
\end{align}

\paragraph{State-space form.}
Let
\[
\mathbf{T}_{g}(t) =
\begin{bmatrix}
T^{\mathrm{jun}}_{g}(t) \\[0.15em]
T^{\mathrm{pack}}_{g}(t) \\[0.15em]
T^{\mathrm{clt}}_{g}(t) \\[0.15em]
T^{\mathrm{rad}}_{g}(t)
\end{bmatrix},
\qquad
u_g(t)=P_{g}(t).
\]
Then the continuous-time dynamics can be written as
\begin{equation}
\frac{d\mathbf{T}_{g}(t)}{dt} = A\mathbf{T}_{g}(t) + B u_g(t) + E\,T_{\mathrm{room}}.
\label{eq:rc-state}
\end{equation}

\paragraph{Matrix definitions.}
\begin{equation}
A =
\begin{bmatrix}
-\frac{1}{C_{\mathrm{jun}} R_{\mathrm{jun,pack}}} &
 \frac{1}{C_{\mathrm{jun}} R_{\mathrm{jun,pack}}} & 0 & 0 \\
 \frac{1}{C_{\mathrm{pack}} R_{\mathrm{jun,pack}}} &
 -\bigl(\frac{1}{C_{\mathrm{pack}} R_{\mathrm{jun,pack}}} + \frac{1}{C_{\mathrm{pack}} R_{\mathrm{pack,clt}}}\bigr) &
 \frac{1}{C_{\mathrm{pack}} R_{\mathrm{pack,clt}}} & 0 \\
 0 &
 \frac{1}{C_{\mathrm{clt}} R_{\mathrm{pack,clt}}} &
 -\bigl(\frac{1}{C_{\mathrm{clt}} R_{\mathrm{pack,clt}}} + \frac{1}{C_{\mathrm{clt}} R_{\mathrm{clt,rad}}}\bigr) &
 \frac{1}{C_{\mathrm{clt}} R_{\mathrm{clt,rad}}} \\
 0 & 0 &
 \frac{1}{C_{\mathrm{rad}} R_{\mathrm{clt,rad}}} &
 -\bigl(\frac{1}{C_{\mathrm{rad}} R_{\mathrm{clt,rad}}} + \frac{1}{C_{\mathrm{rad}} R_{\mathrm{rad,air}}}\bigr)
\end{bmatrix}.
\end{equation}
\[
B =
\begin{bmatrix}
 \frac{1}{C_{\mathrm{jun}}}\\0\\0\\0
\end{bmatrix},
\qquad
E =
\begin{bmatrix}
0\\0\\0\\ \frac{1}{C_{\mathrm{rad}} R_{\mathrm{rad,air}}}
\end{bmatrix}.
\]

\subsection{Discretization and thermal divider ratio}
\label{app:rc-disc-divider}

\paragraph{Discretization.}
With sampling period $\Delta t$, we compute
\[
A_d = \exp(A\Delta t),\quad
B_d = \int_0^{\Delta t}\exp(A\tau)\,d\tau\,B,\quad
E_d = \int_0^{\Delta t}\exp(A\tau)\,d\tau\,E,
\]
yielding the discrete form used in the main text:
\[
\mathbf{T}_{g,j+1} = A_d\mathbf{T}_{g,j} + B_d P_{g,j} + E_d T_{\mathrm{room}}.
\]

\paragraph{Thermal resistance divider ratio.}
\label{app:par-thermak-res}
We define a \emph{thermal resistance divider ratio}
$\mathbf{s}=[1, s^{\mathrm{pack}}, s^{\mathrm{clt}}, s^{\mathrm{rad}}]^\top$ such that, for a target junction
temperature $T^{\mathrm{jun}}$ under boundary $T_{\mathrm{room}}$, the corresponding reference profile is
\[
\mathbf{T}^{\mathrm{ref}}(T^{\mathrm{jun}},T_{\mathrm{room}})
= T_{\mathrm{room}}\mathbf{1} + (T^{\mathrm{jun}}-T_{\mathrm{room}})\mathbf{s}.
\]
For the resistive chain $\mathrm{jun}\!\to\!\mathrm{pack}\!\to\!\mathrm{clt}\!\to\!\mathrm{rad}\!\to\!\mathrm{room}$,
the entries of $\mathbf{s}$ follow from steady heat-flow along the path:
\[
s^{\mathrm{pack}}=\frac{R_{\mathrm{pack,clt}}+R_{\mathrm{clt,rad}}+R_{\mathrm{rad,air}}}{\sum R},\quad
s^{\mathrm{clt}}=\frac{R_{\mathrm{clt,rad}}+R_{\mathrm{rad,air}}}{\sum R},\quad
s^{\mathrm{rad}}=\frac{R_{\mathrm{rad,air}}}{\sum R},
\]
where $\sum R = R_{\mathrm{jun,pack}}+R_{\mathrm{pack,clt}}+R_{\mathrm{clt,rad}}+R_{\mathrm{rad,air}}$.

\subsection{Calibration of RC parameters}
\label{app:rc-calibration}

We parameterize the thermal dynamics of each GPU $g$ with a lumped RC network whose unknown parameters are
$\theta_g=\{R_{\text{jun,pack}},R_{\text{pack,clt}},R_{\text{clt,rad}},R_{\text{rad,air}},C_{\text{jun}},C_{\text{pack}},C_{\text{clt}},C_{\text{rad}}\}$.
We identify $\theta_g$ offline via controlled step-load experiments that excite both heating and cooling transients.

\paragraph{Step-load trace generation.}
Starting from an idle steady state, we program a sequence of constant-power plateaus using NVML/DCGM actuation
(each held for $t_{\text{hold}}$ seconds):
(i) idle baseline ($P\approx P_{\text{idle}}$),
(ii) step-up to a high plateau ($P=P_1$) to drive $T^{\text{jun}}$ upward,
(iii) step-down back to a low plateau ($P=P_0$) to observe the cooling transient.
We repeat this sequence for multiple plateau levels $\{P_1\}$ to improve identifiability.
During each experiment, we log time-aligned traces $\{P_g(t),T^{\text{jun}}_g(t)\}$ at sampling interval $\Delta t$.
When available, we additionally log coolant/radiator proxies (e.g., liquid temperature sensors) to constrain intermediate states.

\paragraph{Parameter fitting.}
Given each logged power trace $P_g(t)$, we simulate the RC network forward in time and estimate $\theta_g$ by minimizing
the junction-temperature trajectory error:
\[
\theta_g^\star
=
\arg\min_{\theta_g}
\sum_{t\in\mathcal{T}}
\left(T^{\text{jun}}_{g,\mathrm{sim}}(t;\theta_g)-T^{\text{jun}}_{g,\mathrm{obs}}(t)\right)^2,
\]
optionally augmented with a weak regularizer to stabilize fitting under partial observability.
To capture configuration differences, we treat $R_{\text{rad,air}}$ as configuration-dependent and re-estimate it for each fan/airflow setting,
while keeping internal resistances and capacitances invariant across configurations for the same GPU.

\paragraph{Online usage.}
After calibration, the heat-budget monitor runs online by integrating the discrete dynamics with live telemetry
(power and $T^{\text{jun}}$) and the ambient setpoint, without requiring external sensors.

\section{Instantiating $\hat{L}_r(a)$ for Common LLM SLO Metrics}
\label{app:slo-mapping}

This appendix shows how the generic latency term $\hat{L}_r(a)$ used in the controller maps to standard LLM serving SLO metrics.
Throughout, $a$ denotes a candidate action at the current job-level decision epoch, and $\mathcal{R}$ denotes the queued requests considered by the controller at that epoch.

\paragraph{Conservative latency bound.}
We first form a conservative latency estimate:
\[
\hat{L}_r(a) = \hat{W}_r(a) + q\!\bigl(f_r(a),\ell_r\bigr)\, b^{\mathrm{pf}}_r,
\]
where $b^{\mathrm{pf}}_r$ is the input prefill length of request $r$ and $\ell_r$ is its prompt-length bucket.
The function $q(f,\ell)$ is a configuration-dependent high-quantile \emph{prefill-time-per-token} profile estimated from traces.
We keep $a$ only if
\begin{equation}
\hat{L}_{\max}(a) \le L_{\max},
\quad\mathrm{where}\quad
\hat{L}_{\max}(a) = \max_{r \in \mathcal{R}} \hat{L}_r(a).
\label{eq:slo-constraints}
\end{equation}
Here $L_{\max}$ denotes the SLO bound. When multiple SLO metrics are used, we apply~\eqref{eq:slo-constraints} to each metric and discard any violated actions.

\paragraph{Queue replay from job-time predictions.}
A candidate action $a$ induces, on each GPU $g$, an ordered job list
\(
a_g = \{j_{g,1},j_{g,2},\dots\}.
\)
HeatiTS provides a predicted service time $\hat{\tau}_{g,j}$ for each job.
For a request $r$ that is assigned to GPU $g=g(r,a)$ and served in job $j_{g,t(r,a)}$ (i.e., it appears at position $t(r,a)$ in $a_g$), we estimate its waiting time by replaying predicted job durations:
\begin{equation}
\hat{W}_r(a) = \sum_{u=1}^{t(r,a)-1} \hat{\tau}_{g,\,j_{g,u}}.
\label{eq:wait-replay}
\end{equation}
The effective frequency level used for request $r$ under action $a$ is the frequency of the job that serves it:
\[
f_r(a) \triangleq f_{g(r,a),\,j_{g(r,a),\,t(r,a)}}.
\]

\paragraph{TTFT.}
Using a configuration-dependent high-quantile prefill-time-per-token profile $q^{\mathrm{pf}}(f,\ell)$, we estimate
\begin{equation}
\widehat{T}^{\mathrm{TTFT}}_r(a)
= \hat{W}_r(a) + q^{\mathrm{pf}}\!\bigl(f_r(a),\ell_r\bigr)\, b^{\mathrm{pf}}_r.
\label{eq:ttft-map}
\end{equation}

\paragraph{TPOT.}
Because the output length is unknown before decoding finishes, TPOT is enforced as a per-token rate constraint.
Using a configuration-dependent high-quantile per-token decode-time profile $q^{\mathrm{dec}}(f)$, we estimate
\begin{equation}
\widehat{T}^{\mathrm{TPOT}}_r(a)
= q^{\mathrm{dec}}\!\bigl(f_r(a)\bigr).
\label{eq:tpot-map}
\end{equation}

\paragraph{Using $\hat{L}_r(a)$ in the controller.}
When the serving stack enforces a metric $M \in \{\mathrm{TTFT},\mathrm{TPOT}\}$ with threshold $T^{M}_{\max}$,
we instantiate
\[
\hat{L}_r(a) \leftarrow \widehat{T}^{M}_r(a),\qquad L_{\max} \leftarrow T^{M}_{\max},
\]
and apply the generic bound $\max_{r\in \mathcal{R}}\hat{L}_r(a)\le L_{\max}$.
If multiple metrics are active, an action is feasible only if it satisfies the bound for each metric.

\section{Limitations and Discussions}
\noindent\textbf{Limitations. } 
HeatCache targets chassis-level AIO liquid-cooled, institution-scale multi-GPU servers, where the coolant loop provides exploitable short-term thermal buffering. We do not design for optimality in facility-plumbed direct-to-chip/CDU systems or immersion cooling, whose thermal states and control knobs differ substantially. 
Our controller operates within a single server and does not coordinate cross-server placement, migration, or multi-tenant fleet policies under shared rack/cluster power caps. 
HeatCache assumes access to standard serving, so it does not require LLM serving model changes, and we do not consider its performance. 
We treat the ambient setpoint as a given warmer setup and focus on IT-side control, rather than co-optimizing HVAC setpoints, CRAC control, airflow management, etc. 
HeatCache relies on a reduced-order lumped RC thermal, which is sufficient for online control; however, it is not intended to capture fine-grained CFD airflow effects or detailed fan and radiator dynamics across arbitrary chassis layouts. 
Accurate headroom estimation is portable in principle, but in practice, it requires a short calibration for each server setup (GPU model and AIO loop configuration, such as coolant volume and flow), and it may need occasional recalibration after hardware aging or maintenance.

\noindent\textbf{Future works. }
In the future, we will extend HeatCache along three axes. 
First, we will generalize calibration and headroom tracking across heterogeneous GPUs and AIO designs by using lightweight self-calibration and online drift detection, reducing per-platform setup effort while remaining safe under aging and maintenance. 
Second, we will scale HeatCache beyond a single server by integrating it with cluster schedulers to coordinate thermal headroom across nodes under shared power and admission constraints. 
Third, we will explore co-design with facility controllers, coupling IT-side thermal-aware scheduling with setpoint and airflow policies to optimize end-to-end energy and reliability, and incorporating carbon- or price-aware objectives for sustainable operation.

\end{document}